# A Comprehensive AI Policy Education Framework for University Teaching and Learning

Cecilia Ka Yuk Chan

Affiliation: The University of Hong Kong

Address: Centre for the Enhancement of Teaching and Learning (CETL), Room CPD-1.81, Centennial Campus, The University of Hong Kong, Pokfulam, Hong Kong

Email: Cecilia.Chan@cetl.hku.hk

Website: https://tlerg.cetl.hku.hk/

**Abstract**

This study aims to develop an AI education policy for higher education by examining the perceptions and implications of text generative AI technologies. Data was collected from 457 students and 180 teachers and staff across various disciplines in Hong Kong universities, using both quantitative and qualitative research methods.

Based on the findings, the study proposes an AI Ecological Education Policy Framework to address the multifaceted implications of AI integration in university teaching and learning. This framework is organized into three dimensions: Pedagogical, Governance, and Operational. The Pedagogical dimension concentrates on using AI to improve teaching and learning outcomes, while the Governance dimension tackles issues related to privacy, security, and accountability. The Operational dimension addresses matters concerning infrastructure and training. The framework fosters a nuanced understanding of the implications of AI integration in academic settings, ensuring that stakeholders are aware of their responsibilities and can take appropriate actions accordingly.

## 1. Introduction

In recent months, there has been a growing concern in the academic settings about the use of text generative artificial intelligence (AI), such as ChatGPT, Bing and the latest, Co-Pilot integrated within the Microsoft Office suite. One of the main concerns is that students may use generative AI tools to cheat or plagiarise their written assignments and exams. In fact, a recent survey of university students found that nearly one in three students had used a form of AI, such as essay-generating software, to complete their coursework (Intelligent.com, 2023). About one-third of college students surveyed (sample size 1000) in the US have utilized the AI chatbot such as ChatGPT to complete written homework assignments, with 60% using the programme on more than half of their assignments. ChatGPT types of generative AI tools is capable of imitating human writing, with some students using it to cheat. The study found that 75% of students believe that using the programme for cheating is wrong but still do it, and nearly 30% believe their professors are unaware of their use of the tool. The study also noted that some professors are considering whether to include ChatGPT in their lessons or join calls to ban it, with 46% of students saying their professors or institutions have banned the tool for homework.

This has led to calls for stricter regulations and penalties for academic misconduct involving AI.

Another concern is that the use of generative AI may lead to a decline in students' writing and critical thinking skills (Civil, 2023), as they become more reliant on automated tools to complete their work. Some academics argue that this could have a negative impact on the quality of education and ultimately harm the students' learning outcomes (Korn & Kelly, 2023; Oliver, 2023; Zhai, 2022).

These concerns have led some universities to ban the use of generative AI in their academic programmes. Eight out of 24 universities in the prestigious UK Russell Group have declared the use of the AI bot for assignments as academic misconduct including Oxford and Cambridge. Meanwhile, many other universities around the world, are rushing to review their plagiarism policies citing concerns about academic integrity (Wood, 2023; Yau & Chan, 2023). Some Australian universities have had to alter their exam and assessment procedures back to pen- and paper-based (Cassidy, 2023; Cavendish, 2023).

However, there are also those who argue that generative AI has the potential to revolutionize education and enhance the learning experience for students. For example, some experts suggest that generative AI could be used to provide personalized feedback and support to students, helping them to identify areas of weakness and improve their skills in an adaptive manner (Kasneci et al, 2023; Sinhaliz, Burdjaco, & Du Preez, 2023).

## 1.1 Generative AI and Generative Pre-trained Transformers

Generative AI is a subset of artificial intelligence (AI) that focuses on creating new data or content rather than analysing and interpreting existing data (McKinsey Consultant, 2023). Generative Pre-trained Transformers (GPT) are a type of generative AI model that use deep learning techniques to generate natural language text. The latest versions of GPT, GPT-3.5 and GPT-4, large language models which are trained on a large corpus of text data, are capable of producing human-like text with high levels of coherence, complexity, and diversity. GPT-3.5 and GPT-4 are both examples of artificial general intelligence (AGI), which is the ability of AI systems to perform any intellectual task that a human can do. Unlike artificial narrow intelligence (ANI), which is designed to perform a specific task, AGI is designed to perform multiple tasks and generalize knowledge across different domains. While GPT-3.5 and GPT-4 are not true AGI systems, they represent significant progress towards achieving AGI by demonstrating the ability to perform a wide range of language tasks and generate human-like text. The development of generative AI models like GPT-3.5 and GPT-4 has the potential to revolutionize many fields, including natural language processing, creative writing, and content generation.

## 1.2 Rationale for an Artificial Intelligence Education Policy

With generative AI tools becoming easily accessible to the public in recent months, they are rapidly being integrated into various fields and industries. This has created an urgent need for universities to develop an AI education policy that prepares students to work with and understand the principles of this technology. There are several rationales supporting this need.

Firstly, AI technology is becoming more prevalent in many sectors of the economy, such as finance, healthcare, and transportation. As a result, graduates will need to have a strong understanding of AI principles in order to succeed in these fields. An AI education policy can provide students with the necessary knowledge and skills to work with AI in a professional capacity.

Secondly, AI has the potential to revolutionize many aspects of society, including education itself. AI can be used to enhance student learning by providing personalized, real-time feedback and adapting to individual learning styles. By educating students on AI, universities can help prepare them to be active participants in the development and implementation of AI technology, ensuring that it benefits society as a whole.

Thirdly, as the use of AI in education and assessment becomes more prevalent, it is essential that students understand the principles behind the technology in order to maintain academic integrity and prevent cheating as mentioned previously. An AI education policy can teach students about the ethical considerations surrounding AI, such as bias and fairness, as well as the potential consequences of using AI in academic contexts.

Fourthly, developing an AI education for university is important to prepare students for the future. AI technology is rapidly advancing, and it is likely to play an increasingly important role in society in the coming years. By providing students and teachers with training in AI, universities can help ensure that graduates are equipped to contribute to the development of AI and to navigate the ethical, social, and economic issues that are likely to arise as AI becomes more widespread. Such training should also help students become competent and responsible users of AI in their daily lives.

Finally, it is also worth mentioning that previous AI policies in education did not anticipate the level of advancements that text-based GPT 3.5 and 4 can now achieved. Given the potential benefits and risks associated with the use of generative AI in education, it is important to develop a proper AI education policy that addresses these concerns and provides guidance on the responsible use of AI.

The study employed a comprehensive approach to data collection, gathering rich quantitative and open-ended survey data from a diverse range of stakeholders in the education community to ensure that it reflects the needs and values of all those involved. The combination of these data sources allowed for a holistic understanding of the topic under investigation, providing a nuanced and multifaceted view of the issues at hand. By doing so, we can help to ensure that the use of generative AI in education is both beneficial and ethical.

### 1.3 Existing Policy on Artificial Intelligence

The aim of this study is to investigate the education policy related to AI, however, it is essential to also scrutinize the existing policies governing AI as a whole. As AI expands its sphere of influence to various sectors in our society, there are increasing concerns over the risks of its usage and how it might impact human activities (AI regulation, 2023; WEF, 2023). Some of the major issues of concern that have drawn the attention of governments around the world include discrimination and bias of AI, loss of privacy and malicious use of AI (Hogenhout, 2021). In view of this, countries have been working on national policies and strategies to provide clearer guidance on AI usage in order to maximize its benefits while mitigating the threats brought by it.

To advocate the responsible and proper management of AI technologies, the centre of focus for most national policies on AI have fallen on the discussion of ethics, which deals with "the standards of right and wrong, acceptable and not acceptable" (Hogenhout, 2021, p.11). Floridi (2021)'s framework for the ethical use of AI, which proposed the 5 core principles of "beneficence, non-maleficence, autonomy, justice and explicability", is referred to by most national policies on AI as a foundation to further develop on. In addition, Dexe & Franke (2020) summarized the AI strategy documents from the Nordic countries and identified various ethical principles as the implicit foundation for further developing policies. The official AI governance framework from Singapore also recognized the "explainable, transparent and fair usage of AI in decision-making process" and "human-centric AI solutions" as the guiding principles of ethical use of AI (IMDA & PDPC, 2020). Apart from individual countries, ethics has been the emphasis of the AI policies published by regional and international bodies. UNESCO developed its guidelines on the ethical use of AI technologies by emphasizing the key idea of human-centeredness and hence, human rights and values laid out in the Universal Declaration of Human Rights (UDHR) are advised to be adopted as the necessary foundation to further promote beneficial and appropriate use of AI technologies (UNESCO, 2021b; 2023). AI strategy in the European Union, as Renda's (2020) analysis pointed out, also focused on ethics and highlighted a human-centric approach to AI. In order to protect EU citizens from the danger of abusive use of advanced technologies, EU proposed its own pillars (legal compliance, ethical alignment and sociotechnical robustness) to ensure the trustworthiness of AI and established a specific AI expert group to work on specific policy recommendations and guidelines.

The heavy focus that these national and regional policies has placed on ethics demonstrates how limited they can do for the implementation of AI technologies. On the one hand, difficulty to lay down a universal definition on ethical principles becomes a hinderance for certain countries in formulating policies on the use of AI (Dexe & Franke, 2020). On the other hand, as AI can weave into the fabrics of everyday human activities, the resulting wide coverage of policy areas ranging from governance to education and even to environment makes it a challenging task for government to establish specific policies on AI usage (UNESCO, 2021b). Thus, as the Singaporean AI governance framework highlighted, model framework or ethical guidelines were in themselves directional and for reference only, and AI practitioners need to consider them with flexibility and according to the relevance of particular situations (IMDA & PDPC, 2020).

Moving forward, the ongoing efforts of national and international organizations to ensure the positive implementation of AI technologies will continue to prioritize discussions and the formulation of legal and ethical principles (AI regulation, 2023; UNESCO, 2023). However, until these principles are validated by real-time implementation of AI technologies, they will remain primarily predictive and prescriptive in nature (Chatterjee, 2020). Over time, it may become necessary for countries to establish institutional support systems to effectively manage AI practices in accordance with validated legal and ethical guidelines (Renda, 2020).

Below consists of a compilation of fundamental ethical principles for AI that have been extracted from multiple policies (IMDA & PDPC, 2020).

| Fundamental Ethical Principles for AI |
| --- |
| 1. Accountability: Ensure AI actors are held responsible for the AI systems' functioning and adherence to ethical principles. |

| | |
|---|---|
| 2. | Accuracy: Recognize and communicate sources of error and uncertainty in algorithms and data to inform mitigation procedures. |
| 3. | Auditability: Allow third parties to examine and review algorithm behavior through transparent information disclosure. |
| 4. | Explainability: Ensure that algorithmic decisions and underlying data can be explained in layman's terms. |
| 5. | Fairness: Prevent discriminatory impacts, include monitoring mechanisms, and consult diverse perspectives during system development. |
| 6. | Human Centricity and Well-being: Prioritize the well-being and needs of humans in AI development and implementation. |
| 7. | Human rights alignment: Ensure technologies do not violate internationally recognized human rights. |
| 8. | Inclusivity: Make AI accessible to everyone. |
| 9. | Progressiveness: Favour projects with significantly greater value than their alternatives. |
| 10. | Responsibility, accountability, and transparency: Build trust through responsibility, accountability, and fairness, provide avenues for redress, and maintain records of design processes. |
| 11. | Robustness and Security: Ensure AI systems are safe, secure, and resistant to tampering or data compromise. |
| 12. | Sustainability: Favour implementations that provide long-lasting, beneficial insights and can predict future behavior. |

Table 1: Compilation of fundamental ethical principles (IMDA &PDPC, 2020)

## 1.4 Existing Policy on AI in Education

The integration of AI technologies into teaching and learning has begun as early as the 1970s and nowadays, different forms of these technologies are used in various educational contexts, such as the use of personalized applications for learning and assessment and information systems that help handle administrative and management tasks in schools (Schiff, 2022; UNESCO, 2021a).

As mentioned above, the use of AI technologies has raised different issues of concern. In the educational contexts, other than the general risks brought by the use of AI, concerns are primarily centred on issues such as what changes can AI bring to the design of assessment and curriculum, equalities and universality in accessing these technologies and the lack of technological infrastructure for emerging economies (Pelletier et al., 2022; TEQSA, 2023; UNESCO, 2021a). Based on these concerns, AI policies in education fix their eyes on addressing a number of issues: literacy education to prevent inequalities in the use of digital technologies (UNESCO, 2021b); essential values of traditional forms of teaching and learning such as teacher-student and student-student relationships (UNESCO, 2021b); inclusiveness and equity in the use of AI technologies (UNESCO, 2021a); and training and enhancement of skills or "micro-credentials" for students that are important and necessary for harnessing technologies (Pelletier et al., 2022; UNESCO, 2021a). The roles of literacy education and skills training are having particular implications for the wider society as the population in general also needs to be prepared for the implementation of AI technologies in different sectors.

Despite identifying multiple issues of concern in the educational contexts, policies on AI in education are mostly generic and implicit because of the lack of concrete evidence of implementing AI technologies (UNESCO, 2021a). In Schiff (2022)'s review on 24 AI policy strategies focusing on the role of education in global AI policy discourse, it was found that policymakers view education largely as an instrumental tool to support workforce development and training of AI experts. The article finds that the use of AI in education is largely absent from policy conversations, while the instrumental value of education in supporting an AI-ready workforce and training more AI experts is overwhelmingly prioritized. The article suggests that if such a trend continues, policymakers may fail to realize AI in education's transformative potential and may fail to sufficiently fund, regulate, and consider AI in education's ethical implications. While more work is still to be done in order to formulate more comprehensive and focused policy documents on AI in education, ethics was reiterated again as a strategically plausible starting point for further discourse and researchers were especially encouraged to engage further with policymakers through their work on ethics in the use of AI in education (Schiff, 2022).

In this research, we will employ the guidelines put forth by UNESCO (UNESCO, 2021a) as the starting point for crafting a more accurate AI policy for university teaching and learning. The rationale for employing UNESCO recommendations as the basis is multifaceted. First, UNESCO is an esteemed international organisation with significant expertise in education, their recommendations are supported by thorough research and knowledge from experts worldwide. These recommendations are designed to be relevant and flexible for a variety of educational systems and cultural settings, making them suitable for diverse institutions. UNESCO's guidelines also take a comprehensive approach to incorporating AI in education, addressing important ethical, social, economic, and technological aspects essential for creating effective policies. Using an existing framework like UNESCO's recommendations saves time and resources and provides a well-organized starting point for examining specific AI policy issues in university teaching and learning. Finally, anchoring the study in UNESCO's recommendations enhances the credibility of the research.

The UNESCO framework for AI in education is centred around a humanistic approach, which aims to safeguard human rights and provide individuals with the necessary skills and values for sustainable development, as well as effective human-machine collaboration in life, learning, and work. The framework prioritizes human control over AI and ensures that it is utilized to improve the capabilities of both teachers and students. Moreover, the framework calls for ethical, transparent, non-discriminatory, and auditable design of AI applications. From the UNESCO's AI and Education: Guidance for Policy-Makers document, the following recommendations are provided:

1. Interdisciplinary planning and inter-sectoral governance: This recommendation suggests that AI and education policies should be developed through collaboration between different sectors and disciplines to ensure a comprehensive approach. For example, policymakers could work with experts in education, technology, ethics, and other relevant fields to develop policies that take into account all aspects of AI in education.
2. Policies on equitable, inclusive, and ethical use of AI: This recommendation emphasizes the importance of ensuring that AI is used in an ethical and inclusive manner that benefits all learners. For example, policymakers could develop policies that address issues such as bias in AI algorithms or access to AI tools for learners from disadvantaged backgrounds.
3. Develop a master plan for using AI for education management, teaching, learning, and assessment: This recommendation suggests that policymakers should develop a comprehensive plan for using AI in various aspects of education to ensure its effective implementation. For example, a master plan could include specific goals for using AI in areas such as personalized learning or teacher professional development.

4. Pilot testing, monitoring and evaluation, and building an evidence base: This recommendation highlights the importance of testing and evaluating the use of AI in education through pilot projects to build an evidence base for its effectiveness. For example, policymakers could fund pilot projects that test the use of AI tools in specific educational contexts or with specific learner populations.
5. Fostering local AI innovations for education: This recommendation suggests that policymakers should encourage the development of local innovations in AI for education to ensure that it meets the specific needs of their communities. For example, policymakers could provide funding or support to local startups or research institutions working on developing new AI tools or applications specifically designed for their region's educational needs.

This study aims to use UNESCO's recommendations as a basis for developing AI education policy framework for university teaching and learning and will gather input from various stakeholders to identify any gaps in the framework. Based on their ideas, recommendations, and concerns, the framework will be modified and adapted accordingly.

## 2. Methodology

In this study, a survey design was utilized to gather data from students, teachers, and staff in Hong Kong to examine their usage and perceptions of generative AI in teaching and learning. The survey was administered through an online questionnaire, featuring a mix of closed-ended and open-ended questions. Topics covered in the survey included the use of generative AI technologies like ChatGPT, the integration of AI technologies in higher education, potential risks associated with AI technologies, and AI's impact on teaching and learning.

Data were collected via an online survey from a diverse group of stakeholders in the education community, ensuring that the results reflect the needs and values of all participants. A convenience sampling method was employed for selecting the respondents, based on their availability and willingness to participate in the study. Participants were recruited through an online platform and provided with an informed consent form prior to completing the survey.

The survey was completed by 457 undergraduate and postgraduate students, as well as 180 teachers and staff members across various disciplines in Hong Kong. Descriptive analysis was used to analyse the survey data, while a thematic analysis approach was applied to examine the responses from the open-ended questions in the survey.

### 2.1 Quantitative Data (Survey Data) and Descriptive Analysis:

A range of survey items was included to capture different aspects of participants' usage and perception of generative AI technologies like ChatGPT. For example, participants were asked whether they have used ChatGPT or similar generative AI technologies before and how they envision using these technologies in their teaching and learning practices.

Descriptive analysis was employed to analyse the survey data collected from students and teachers in Hong Kong, in order to gain a better understanding of the usage and perception of generative AI technologies like ChatGPT in higher education. Descriptive analysis is an appropriate statistical method for summarizing and describing the main characteristics of the sample and the data collected. It is particularly useful for analysing survey data and can provide an overview of the distribution, central tendency, and variability of the responses.

## 2.2 Qualitative Data (Open-ended Data) and Thematic Analysis:

Aside from the quantitative part of the survey, respondents were also asked about their apprehensions regarding the integration of generative AI technologies and their recommendations for university strategic plans through open-ended questions. Different perspectives and experiences were gathered from the respondents across various disciplines. The data from the open-ended questions were analysed using a thematic analysis approach, which involved identifying patterns and themes in the data. An inductive approach was used to analyse the responses, where the themes emerged from the data rather than being predetermined by the researcher.

The combination of the quantitative and qualitative data enabled a more holistic understanding of the usage and perception of generative AI technologies in higher education. This allowed for the pinpointing of potential requirements, recommendations, and strategies for AI policy in university teaching and learning. This understanding is essential for ensuring that the use of these technologies is both beneficial and ethical.

## 3. Results
## 3.1 Findings from the Quantitative Data

The survey was conducted among 457 students and 180 teachers and staff from different disciplines in Hong Kong universities. The goal was to explore the kinds of requirements, guidelines and strategies necessary for developing AI policies geared towards university teaching and learning. The findings reveal valuable insights into the perception of generative AI technologies like ChatGPT among students and teachers (refer to table 2).

Regarding the usage of generative AI technologies, both students (mean=2.28, SD=1.18) and teachers (mean=2.02, SD=1.1) reported relatively low experience, suggesting that there is significant room for growth in adoption. Both groups demonstrated a belief in the positive impact of integrating AI technologies into higher education (students: mean=4, SD=0.891; teachers: mean=3.87, SD=1.32). This optimism was also reflected in the strong agreement that institutions should have plans in place associated with AI technologies (students: mean=4.5, SD=0.854; teachers: mean=4.54, SD=0.874).

Both students and teachers were open to integrating AI technologies into their future teaching and learning practices (students: mean=3.93, SD=1.09; teachers: mean=3.92, SD=1.31). However, there were concerns among both groups about other students using AI technologies to get ahead in their assignments (students: mean=3.67, SD=1.22; teachers: mean=3.93, SD=1.12). Interestingly, both students and teachers did not strongly agree that AI technologies would replace teachers in the future (students: mean=2.14, SD=1.12; teachers: mean=2.26, SD=1.34).

The respondents acknowledged the importance of learning to use generative AI technologies well for their careers (students: mean=4.07, SD=0.998; teachers: mean=4.1, SD=1.08). However, both groups expressed doubt about teachers' ability to accurately identify a student's usage of generative AI technologies for completing assignments (students: mean=3.02, SD=1.56; teachers: mean=2.72, SD=1.62).

The responses to the remaining questions suggest that students and teachers recognize potential benefits and drawbacks of AI technologies, including providing guidance and personalized feedback, improving digital competence and academic performance, and offering anonymity in student support services. However, there were concerns about over-reliance on AI technologies, limited social interaction, and the potential hindrance to the development of generic skills.

These findings highlight the need for a comprehensive AI policy in higher education that addresses the potential risks and opportunities associated with generative AI technologies. Based on these findings, some implications and suggestions for university teaching and learning AI policy include:

1. Training: Providing training for both students and teachers on effectively using and integrating generative AI technologies into teaching and learning practices.
2. Ethical Use and Risk Management: Developing policies and guidelines for ethical use and risk management associated with generative AI technologies.
3. Incorporating AI without replacing human: Incorporating AI technologies as supplementary tools to assist teachers and students, rather than as replacements for human interaction.
4. Continuously Enhancing Holistic Competencies: Encouraging the use of AI technologies to enhance specific skills, such as digital competence and time management, while ensuring that students continue to develop vital transferable skills.
5. Fostering a transparent AI environment: Fostering a transparent environment where students and teachers can openly discuss the benefits and concerns associated with using AI technologies in higher education.
6. Data Privacy and security: Ensuring data privacy and security while using AI technologies.

Overall, the survey results indicate an openness to adopting generative AI technologies in higher education and a recognition of the potential advantages and challenges. Addressing these issues through informed policy and institutional support will be crucial for maximizing the benefits of AI technologies in university teaching and learning.

| Item | Students | | | | Teachers | | | |
|---|---|---|---|---|---|---|---|---|
| | N | Mean | Median | SD | N | Mean | Median | SD |
| I have used generative AI technologies like ChatGPT. | 457 | 2.28 | 2 | 1.18 | 180 | 2.02 | 2 | 1.1 |
| The integration of generative AI technologies like ChatGPT in higher education will have a positive impact on teaching and learning in the long run. | 457 | 4 | 4 | 0.891 | 180 | 3.87 | 4 | 1.32 |
| Higher education institutions should have a plan in place for managing the potential risks associated with using generative AI technologies like ChatGPT in teaching and learning. | 457 | 4.5 | 5 | 0.854 | 180 | 4.54 | 5 | 0.874 |
| I envision integrating generative AI technologies like ChatGPT into my teaching and learning practices in the future. | 455 | 3.93 | 4 | 1.09 | 180 | 3.92 | 4 | 1.31 |
| I am concerned that other students may use generative AI technologies like ChatGPT to get ahead in their assignments. /I am concerned that there may be an unfair advantage for some students as they may use generative AI technologies like ChatGPT to get ahead in their assignments. | 456 | 3.67 | 4 | 1.22 | 180 | 3.93 | 4 | 1.12 |
| AI technologies like ChatGPT will replace teachers in the future. | 457 | 2.14 | 2 | 1.12 | 180 | 2.26 | 2 | 1.34 |
| Students must learn how to use generative AI technologies well for their career. | 457 | 4.07 | 4 | 0.998 | 180 | 4.1 | 4 | 1.08 |
| Teachers can already accurately identify a student's usage of generative AI technologies to partially complete an assignment. | 457 | 3.02 | 3 | 1.56 | 180 | 2.72 | 2 | 1.62 |
| Generative AI technologies such as ChatGPT can provide guidance for coursework as effectively as human teachers. | 455 | 3.19 | 3 | 1.25 | 180 | 2.93 | 3 | 1.4 |

| Statement | N | Mean | Median | SD | N | Mean | Median | SD |
|---|---|---|---|---|---|---|---|---|
| Using generative AI technologies such as ChatGPT to complete assignments undermines the value of a university education. | 455 | 3.29 | 3 | 1.25 | 180 | 3.56 | 4 | 1.31 |
| I can ask questions to generative AI technologies such as ChatGPT that I would otherwise not voice out to my teacher. /Students can ask questions to generative AI technologies such as ChatGPT that they would otherwise not voice out to their teacher. | 454 | 3.51 | 4 | 1.2 | 180 | 3.97 | 4 | 1.06 |
| Generative AI technologies such as ChatGPT will not judge me, so I feel comfortable with it. /Students will not feel judged by generative AI technologies such as ChatGPT, so they feel comfortable with it. | 452 | 3.66 | 4 | 1.15 | 180 | 4 | 4 | 1.17 |
| Generative AI technologies such as ChatGPT will limit my opportunities to interact with others and socialize while completing coursework. /Generative AI technologies such as ChatGPT will limit students' opportunities to interact with others and socialize while completing coursework. | 454 | 3.24 | 3 | 1.32 | 180 | 3.69 | 4 | 1.3 |
| Generative AI technologies such as ChatGPT will hinder my development of generic or transferable skills such as teamwork, problem-solving, and leadership skills. /Generative AI technologies such as ChatGPT will hinder students' development of generic or transferable skills such as teamwork, problem-solving, and leadership skills. | 454 | 3.3 | 3 | 1.33 | 180 | 3.74 | 4 | 1.41 |
| If a fully online programme with the assistance of a personalized AI tutor was available, I would be willing to pursue my degree through this option. /If a fully online programme with the assistance of a personalized AI tutor was available, students should be open to pursuing their degree through this option. | 454 | 2.92 | 3 | 1.46 | 180 | 3.21 | 3 | 1.52 |
| I can become over-reliant on generative AI technologies. /Students can become over-reliant on generative AI technologies. | 454 | 3.11 | 3 | 1.35 | 180 | 4.24 | 4 | 0.955 |
| I believe generative AI technologies such as ChatGPT can improve my digital competence. /I believe Generative AI technologies such as ChatGPT can improve students' digital competence. | 454 | 3.8 | 4 | 1.06 | 180 | 3.83 | 4 | 1.12 |
| I believe generative AI technologies such as ChatGPT can improve my overall academic performance. /I believe Generative AI technologies such as ChatGPT can improve students' overall academic performance. | 455 | 3.67 | 4 | 1.18 | 180 | 3.63 | 4 | 1.36 |
| I believe generative AI technologies such as ChatGPT can help me save time. /I believe Generative AI technologies such as ChatGPT can help students save time. | 453 | 4.23 | 4 | 0.848 | 180 | 4.06 | 4 | 1.01 |
| I think generative AI technologies such as ChatGPT can help me become a better writer. /I think Generative AI technologies such as ChatGPT can help students become a better writer. | 455 | 3.46 | 4 | 1.27 | 180 | 3.31 | 3 | 1.45 |
| I believe AI technologies such as ChatGPT can provide me with unique insights and perspectives that I may not have thought of myself. /I believe AI technologies such as ChatGPT can provide students with unique insights and perspectives that they may not have thought of themselves. | 455 | 3.84 | 4 | 1.13 | 180 | 3.77 | 4 | 1.26 |
| I think AI technologies such as ChatGPT can provide me with personalized and immediate feedback and suggestions for my assignments. /AI technologies such as ChatGPT can provide students with personalized and immediate feedback and suggestions for their assignments. | 455 | 3.75 | 4 | 1.14 | 180 | 3.86 | 4 | 1.34 |
| I think AI technologies such as ChatGPT is a great tool as it is available 24/7. /I think AI technologies such as ChatGPT is a great tool for students as it is available 24/7. | 455 | 4.16 | 4 | 0.893 | 180 | 3.81 | 4 | 1.17 |
| I think AI technologies such as ChatGPT is a great tool for student support services due to anonymity. | 455 | 3.91 | 4 | 1.12 | 180 | 3.77 | 4 | 1.29 |

Table 2: Descriptive analysis for Quantitative Results

### 3.2 Findings from the Qualitative Data

The qualitative data collected from students, teachers, and staff yielded valuable and rich suggestions and comments. From the data, we identified ten key areas that are directly relevant to the planning of an AI policy for teaching and learning in universities. These areas align well with the quantitative data and are as follows:

**(1) Understanding, Identifying and Preventing Academic Misconduct and Ethical Dilemmas**

To address academic misconduct, universities must develop clear guidelines and strategies for detecting and preventing the misuse of generative AI. Teachers emphasize the importance of creating university-wide policies on how to test students suspected of using AI to complete tasks in which AI use is prohibited or misused. As one student stated, "*A clear set of rules about what happens if AI is used and resources on informing students about the rule set are needed.*" They also suggested, "*Clearly stipulate in which areas generative AI technologies are allowed and which are not. What are the procedures to handle suspended cases? What are the consequences?*" Another student mentioned that "*the level of restriction should be clarified.*" Both teachers and students have also suggested the use of assessments that minimize opportunities for AI misuse, such as oral examinations or controlled settings where internet access is limited, to help maintain academic integrity. Both teachers and students have also questioned "*what is the definition of cheating?*" in this AI era.

Teachers highlight the importance of identifying ethical dilemmas and recommend familiarizing students with ethical issues, such as the boundaries between plagiarism and inspiration and appropriate situations for seeking help from AI. Establishing clear policies around AI use, including ethical guidelines and legal responsibilities, will help students and staff navigate these complex issues. One teacher noted, "*The education on academic and research ethics should be strengthened*." Explicitly stipulating the areas where AI is allowed and the procedures for handling suspected cases of misuse will help maintain a transparent and equitable learning environment.

**(2) Addressing Governance of AI: Data Privacy, Transparency, Accountability and Security**

Universities must take responsibility for decisions made regarding the use of generative AI in teaching and learning, which includes being transparent about data collection and usage, and being receptive to feedback and criticism. By disclosing information about the implementation of generative AI, including the algorithms employed, their functions, and any potential biases or limitations, universities can foster trust and confidence among students and staff in AI technology usage. Teachers emphasize the importance of addressing ethical concerns, privacy, security, and other related issues when using generative AI technologies. Teachers commented "*In general, its impact is inevitable. It may negatively affect social consciousness and responsibility. Depending on climate change management and its consequences, it may contribute to the demise of a significant portion of humanity. It may also protect and advance the interests of those who benefit from chaos.*"

Privacy and Security: AI technologies rely on vast amounts of data, which raises concerns about privacy and security if the data is not adequately protected. "*Institutions should ensure that the data used by generative AI technologies is kept private and secure. This includes ensuring that any data used in training or testing the technology is de-identified, and that appropriate security measures are in place to prevent unauthorized access or use of data.*"

Transparency and Accountability: Universities should be transparent about the use of generative AI in teaching and learning, which includes disclosing information about the algorithms and their functions, as well as any potential biases or limitations of the AI tools. "*It is essential to recognize ethical dilemmas and consider privacy, security, and related issues when employing generative AI technologies.*"

The complexity of AI technologies can make it difficult to hold organizations and individuals accountable for their decisions and actions. Institutions should address ethical issues, such as potential discrimination, bias, and stereotypes, while ensuring data privacy and security.

### (3) Monitoring and Evaluating AI Implementation

To ensure the success of AI integration in university teaching and learning, continuous monitoring and evaluation of its implementation are necessary. Teachers recommend conducting longitudinal experiments in different areas to better understand how AI affects students' learning processes and outcomes. Regular assessments of AI's impact on teaching practices and student performance will help identify areas for improvement and ensure that the technology is being used effectively and ethically. One student mentioned, "*The plan should include more experiments on conducting the AI technologies on teaching.*" By regularly collecting feedback from both teachers and students, universities can make informed decisions about how to improve AI implementation. Evaluating the effectiveness of AI tools in enhancing learning outcomes is vital in determining their value and making adjustments as needed.

### (4) Ensuring Equity in Access to AI Technologies

Ensuring equitable access to AI technologies is crucial for fostering an inclusive learning environment. Universities should work to provide resources and support to all students and staff, regardless of their background or access to technology. This may involve the procurement of AI tools, including AI detectors, for use by the entire university community. By promoting equal access to AI technologies, universities can help level the playing field and ensure that all students and staff have the opportunity to benefit from the advantages offered by AI integration.

Equal access to AI technologies is essential for maintaining fairness in the educational environment. One teacher commented, "*Same as all other resources, to incorporate this into current industries (especially education), fairness should be a top priority. If the usage involves any kind of competition, e.g. access to ChatGPT should be equal for all involved parties.*" Another student highlights "*Ethical dilemma includes ensuring that the technology is not used to discriminate against individuals or groups, and that it does not reinforce bias or stereotypes.*" Universities should consider how to ensure that all students have access and training to AI tools and resources, regardless of their socio-economic backgrounds, in order to level the playing field and promote inclusivity.

### (5) Attributing AI technologies

Attribution is an important aspect of AI policy in university teaching and learning. One student remarked, "*They are welcome to use AI for academic purposes while requiring students to state clearly which part was helped by AI. This is similar to the references and citation of current academic practice.*" By requiring students to attribute AI-generated content, universities can promote academic integrity and ensure that AI technologies are used ethically in the learning process. Furthermore, there is a need for guidelines on how to fairly attribute generative AI's contribution to student work. "*Ethics of use, knowledge of affordances, effective use, critique/evaluation of outputs, and role/integration in workflows/product in study and professional settings*" may be included in the attribution.

**(6) Providing Training and Support for Teachers, Staff and Students in AI Literacy**

To ensure successful integration of AI in teaching and learning, universities must provide adequate training and support for teachers, staff, and students. Teachers express concerns about coping with this new trend, helping students use AI effectively, and learning from student usage. As one teacher puts it, "*Staff and students need an educative approach to its ethical use.*" Investing in training and resources can help educators feel more confident and capable in navigating the complexities of AI in their classrooms. This is supported by many students and teachers who believe that institutions should provide training to faculty and staff on the appropriate use of generative AI technologies in teaching and learning. "*This training should include information on selecting appropriate technologies, using them effectively, and managing the risks associated with their use,*" they say. In addition, "*teaching students how to use the technology and how to critique it is probably central to successfully planning for the integration of AI in education.*" Students suggest that "*teaching students the potential of using generative AI properly and critically can benefit from students using AI hiddenly*" and "*relying on tools able to detect the use of a generative language model, while being aware of the limits of such tools.*"

AI literacy is crucial for both students and staff as they navigate the use of generative AI in teaching and learning. Teachers emphasize the need for education on ethics, knowledge of AI tool affordances, effective use (e.g., prompt engineering), critique and evaluation of outputs, and the role of AI in study and professional settings. A comprehensive AI literacy programme will help students and staff better understand and responsibly utilize AI technologies in their academic and professional lives. By providing training and resources on AI technologies, universities can empower students and staff to make informed decisions about their use and potential applications in teaching and learning.

**(7) Rethinking Assessments and Examinations**

The integration of generative AI in education calls for a re-evaluation of assessments and examinations. Teachers suggest designing assessments that allow AI technologies to enhance learning outcomes, rather than solely producing outputs. For example, one teacher recommends "*Promote assessments and activities where students can by themselves discover the limits of such techniques - and relativize the idea that they could be useful to 'cheat'.*" This shift may necessitate the development of new assessment methods that balance the benefits of AI with the need to maintain academic integrity. A student stated, "*Change of assessment methods to measure the true 'understanding' of students instead of the ability to collect information (which can easily be done with AI tools).*" Universities may need to develop new assessment strategies that focus on students' understanding, critical thinking, and analysis to prevent AI-generated content from compromising the assessment process. A teacher noted that "*... it is hard to assess most of them, so we fall back on regurgitation*", a change is necessary.

**(8) Encouraging a Balanced Approach to AI Adoption**

A balanced approach to AI adoption in university teaching and learning involves recognizing both the potential benefits and limitations of generative AI technologies. One teacher suggests,

"*Be positive about this technological evolution and incorporate it to develop new assignments and assessment.*" This approach requires flexibility, striking a balance between embracing new technology for its potential to enhance efficiency and productivity while maintaining a focus on critical thinking and ethical considerations. It is also important to encourage a balanced approach to AI adoption to avoid over-reliance on these technologies. "*We should learn how AI can assist us, but not replace schoolwork,*" one teacher advised. This approach involves using AI technologies as complementary tools to support learning rather than relying on them as a substitute for traditional teaching methods. Students should be encouraged to use AI as an aid to their learning process and not solely depend on it for academic success.

### (9) Preparing Students for the AI-Driven Workplace

Preparing students for an AI-driven workplace involves teaching them how to use AI responsibly, ethically, and effectively. Universities should develop curricula that reflect the increasing prominence of AI in various industries, ensuring that students are equipped with the skills and knowledge to navigate the evolving workplace landscape. This includes teaching students how to integrate AI into their workflows, evaluate the effectiveness of AI tools, and understand their role in professional settings. As one teacher notes, "*Teaching students how to use it properly and understanding its limitations and strengths would be useful.*"

Integrating AI technologies into teaching and learning involves familiarizing students with AI tools they will likely encounter during their university studies and in the workplace, as mentioned by a student who said, "*Teach students how to best use AI tools and make AI tools a common part of education, just like PowerPoint and Excel.*" Teachers suggest guiding students to recognize ethical issues and helping them self-appropriate AI in study and work settings.

As the workplace increasingly adopts AI technologies, universities should prepare students for this shift. One student stated, "*Plans should be implemented to assist students in making better and more constructive use of AI in learning, career planning, and personal development.*"

### (10) Developing Student Holistic Competencies/Generic Skills

Teachers have highlighted the importance of enhancing critical thinking, digital literacy, information literacy, and professional ethics among students to help them make effective and ethical use of AI technologies. To harness the potential of generative AI technologies, teachers advocate for an emphasis on teaching students to assess the reliability of content, understand biases, and evaluate the accuracy and relevance of AI-generated information. One teacher suggests that "*Enhancement on critical thinking among students is definitely a must, in order to make good use of such AI technologies.*" Another teacher emphasizes the importance of "*Teaching students how to use the technology, and how to critique it, is probably central to successfully planning for the integration of AI in education.*"

To successfully embrace generative AI technologies, universities should prioritize fostering critical thinking among students. One student suggested, "*Plans on how to maintain students' interest and motivation to engage in deep and critical thinking, diversify perspectives and expand horizons.*"

Developing holistic competencies and generic skills in students is an essential goal of education. "*However there are many things, even far more important things, to education…: the education*

*of character; rhetoric and analytical skills; public speaking; creativity; memorisation; all embodied skills*," a teacher argued. Incorporating AI technologies into teaching and learning may hinder students' development of competencies such as teamwork, leadership, empathy, and creativity skills. Therefore, universities need to continuously find opportunities for students to develop these skills, preparing them for the AI-driven workplace where they need to be adaptable, resilient, and transformational.

## 4. Discussion
### 4.1 Triangulating Quantitative and Qualitative Data

The quantitative findings support the key areas found in the qualitative data for AI integration in education. The quantitative data reveals that both students and teachers share concerns about the potential misuse of AI technologies, such as ChatGPT, in assignments (students: mean 3.67, teachers: mean 3.93). This emphasizes the need for guidelines and strategies to prevent academic misconduct. Furthermore, there is significant agreement among students and teachers on the necessity for higher education institutions to implement a plan for managing the potential risks associated with using generative AI technologies (students: mean 4.5, teachers: mean 4.54), highlighting the importance of addressing data privacy, transparency, accountability, and security. The overall positive perception of AI technologies integration within education implies that proper policies should be in place to ensure responsible AI incorporation in higher education.

The concern that some students might use generative AI technologies to gain an advantage in their assignments (students: mean 3.67, teachers: mean 3.93) underscores the importance of ensuring equal access to AI technologies for all students. Moreover, the consensus that students must become proficient in using generative AI technologies for their careers (students: mean 4.07, teachers: mean 4.1) highlights the need for AI literacy and training for all stakeholders in the educational process, preparing students for the AI-driven workplace.

Interestingly, teachers and students are unsure if teachers can accurately identify a student's use of generative AI technologies to partially complete an assignment (students: mean 3.02, teachers: mean 2.72), yet they also believe that AI technologies can provide unique insights and perspectives and personalized feedback. This suggests that rethinking assessment methods may be necessary.

Data indicating that neither students nor teachers believe AI technologies will replace teachers in the future (students: mean 2.14, teachers: mean 2.26) supports the need for a balanced approach to AI adoption, utilizing AI technologies as complementary tools rather than substitutes for traditional teaching methods. Finally, concerns that generative AI technologies could hinder students' development of generic or transferable skills, such as teamwork, problem-solving, and leadership (students: mean 3.3, teachers: mean 3.74), emphasize the importance of focusing on students' holistic competencies and generic skills in preparation for the AI-driven workplace.

### 4.2 Key Areas versus UNESCO's Recommendations on AI Education Policy

The original plan for the study was to use UNESCO's recommendations as a basis for developing AI education policy framework for university teaching and learning through inputs from various stakeholders to identify any gaps in the framework and modified accordingly. Although the recommendations from UNESCO can provide a high-level guideline for this study, it was clear that there are several key differences between UNESCO recommendations and the ten key areas that were identified for integrating AI in university teaching and learning.

As the UNESCO's AI and Education: Guidance for Policy-Makers was written before the availability of GPT 3.5 and 4, the recommendations would not have fully addressed the current opportunities and threats of the advances in the GPT technologies for education.

Moreover, the UNESCO recommendations are intended for education in general and do not specifically cater to the needs of university teaching and learning. The UNESCO recommendations are high-level and general aimed at helping policy-makers better understand the possibilities and implications of AI for teaching and learning to help achieve Sustainable Development Goal 4: Ensure inclusive and equitable quality education and promote lifelong learning opportunities for all (UNESCO, 2021a), while the ten key areas are more specific, practical, and tailored to university teaching and learning.

The ten key areas were developed based on direct input from stakeholders who have vested interests in university teaching and learning, which makes them more relevant and grounded in practice. For example, UNESCO's recommendation to develop a master plan for using AI for education management, teaching, learning, and assessment emphasizes the need for a comprehensive and strategic approach to integrating AI in various aspects of education. This includes not only teaching and learning but also the broader aspects of education management, such as administration, resource allocation, and policy development. The focus here is on creating an overarching framework that guides the implementation of AI in education as a whole. On the other hand, the key area derived from the qualitative findings of rethinking assessments and examinations delves deeper into a specific aspect of education: the evaluation of students' learning. This area acknowledges that the integration of generative AI in education necessitates a re-evaluation of traditional assessment methods. The focus here is on designing assessments that allow AI technologies to enhance learning outcomes while maintaining academic integrity. This involves developing new assessment strategies that focus on students' understanding, critical thinking, and analysis rather than just their ability to collect information.

In essence, the difference between these two areas lies in their scope and focus. UNESCO's recommendation is broader, encompassing various aspects of education and advocating for a comprehensive master plan. In contrast, the key area on rethinking assessments and examinations is more specific and targeted, addressing the challenges and opportunities associated with AI integration in student evaluation.

### 4.3 AI Ecological Education Policy Framework

In order to turn policy recommendations into action plans, the ten key areas have been further organised into three dimensions – Pedagogical, Ethical, and Operational into the AI Ecological Education Policy Framework. Each dimension is led by a responsible party (see figure 1). This framework allows for a more nuanced understanding of the multifaceted implications of AI integration in university settings and ensures that stakeholders consider the broader context of AI adoption and its impact on various aspects of teaching and learning.

**Pedagogical Dimension (Teachers)**

This dimension focuses on the teaching and learning aspects of AI integration. It includes the following key areas:

a. Rethinking assessments and examinations
   b. Developing student holistic competencies/generic skills
   c. Preparing students for the AI-driven workplace
   d. Encouraging a balanced approach to AI adoption

**Governance Dimension (Senior Management)**

This dimension emphasizes the governance considerations surrounding AI usage in education. It encompasses the following key areas:

   a. Understanding, identifying, and preventing academic misconduct and ethical dilemmas
   b. Addressing governance of AI: data privacy, transparency, accountability, and security
   c. Attributing AI technologies
   d. Ensuring equity in access to AI technologies

**Operational Dimension (Teaching and Learning and IT staff)**

This dimension concentrates on the practical implementation of AI in university settings. It includes the following key areas:

   a. Monitoring and evaluating AI implementation
   b. Providing training and support for teachers, staff, and students in AI literacy

For the Pedagogical dimension, the framework emphasizes the need to adapt teaching methods and assessment strategies in response to AI's growing capabilities, preparing students for an increasingly AI-driven workplace. By focusing on pedagogy, the framework ensures that AI technologies are harnessed to enhance learning outcomes and develop critical thinking, creativity, and other essential skills, rather than undermining academic integrity. Teachers are the initiator for the Pedagogical Dimension, as they are the ones who design and implement lesson plans, activities, and assessments that utilize AI technologies. They will need to have the expertise to determine how AI can best support and enhance students' learning experiences. By assigning teachers the responsibility for this dimension, we ensure that AI tools are used in a way that is pedagogically sound and enhances the learning outcomes of students.

The Governance dimension highlights the importance of addressing issues related to academic misconduct, data privacy, transparency, and accountability. The framework ensures that stakeholders understand and address the ethical challenges associated with AI technologies, fostering responsible use and helping to maintain trust within the university community. This focus on governance encourages universities to develop clear policies and guidelines, ensuring that students and staff can navigate the complex ethical landscape surrounding AI.

Senior management will be the initiator for the Governance Dimension of the AI Ecology Framework. As they hold decision-making authority, they are tasked with developing and enforcing policies, guidelines, and procedures that address the ethical concerns surrounding AI use in education. These include academic integrity, data privacy, transparency, accountability, and security. Senior management's role is to ensure that AI is used responsibly and ethically, fostering a learning environment that is fair, equitable, and inclusive.

The Operational dimension of the framework underscores the need for ongoing monitoring, evaluation, and support to ensure the effective and equitable implementation of AI technologies. By considering operational aspects, the framework encourages universities to provide training, resources, and support to all stakeholders, promoting equal access to AI technologies and fostering an inclusive learning environment. Furthermore, the operational dimension emphasizes the importance of continuous improvement and adaptation, enabling universities to refine their AI integration strategies in response to new insights and changing needs.

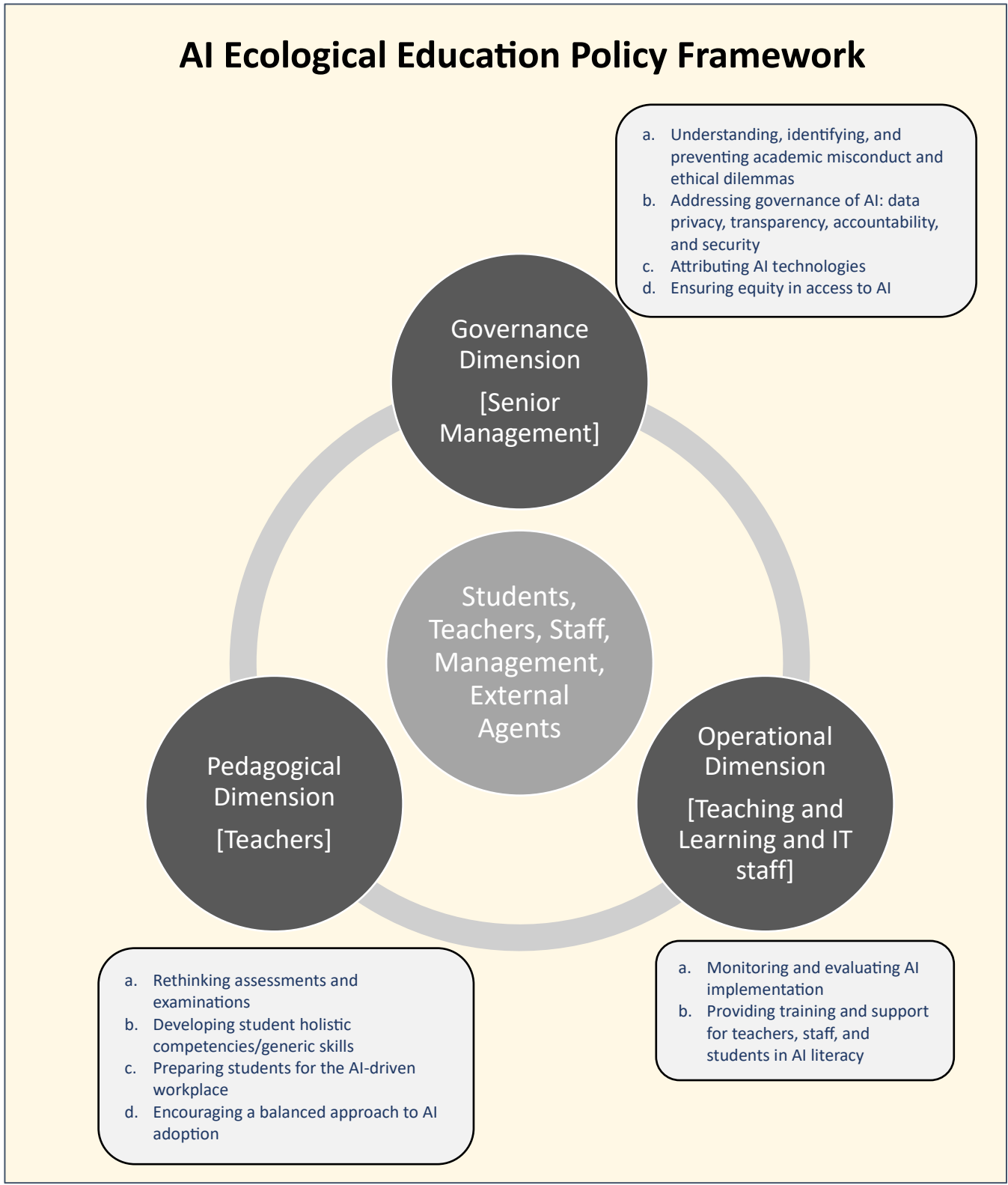

Figure 1: AI Ecological Education Policy Framework

Teaching and Learning and IT staff will be tasked to look after the Operational Dimension. They play a crucial role in managing and maintaining the AI technologies used in the educational setting. Their tasks include providing training and support for both students and staff, ensuring the proper functioning of AI tools, and addressing any technical issues that may arise. They can ensure that AI technologies are seamlessly integrated into the educational environment, minimizing disruptions and maximizing their potential benefits.

**It is crucial to recognize that the responsibility of each dimension in the ecological framework should not be viewed in isolation. Collaboration and communication among all stakeholders (universities, teachers, students, staff and external agents such as accreditation, quality assurance bodies) are essential to ensure the successful implementation of any policy. Each group should actively participate in the development and execution of AI-related initiatives and work together to achieve the desired outcomes in university teaching and learning.**

## 5. Conclusions

This study aims to establish an AI education policy for university teaching and learning, addressing concerns related to the use of text-generating AI in academic environments, such as cheating and plagiarism. It offers valuable insights into the perceptions of generative AI technologies in higher education based on quantitative and qualitative data collected from various stakeholders. However, this study has some limitations, including a relatively small sample size that may not be representative of all educational institutions. Additionally, the research only focused on text-based generative AI technology and did not explore other types or variations. Lastly, the study relied on self-reported data from participants, which may be subject to bias or inaccuracies.

Despite these limitations, this study proposes an AI Ecological Education Policy Framework to address the diverse implications of AI integration in university settings. The framework consists of three dimensions - Pedagogical, Governance, and Operational - each led by a responsible party. This structure allows for a more comprehensive understanding of AI integration implications in teaching and learning settings and ensures stakeholders are aware of their responsibilities. In general, all dimensions as shown in figure 1 are collaborated among different stakeholders, in particular, students need to play an active role in the drafting and the implementation of the policy, that is also the rationale, why students' perception are collected in this study. It is worth mentioning students in this era have been immersed in digital competency since their toddler years, and their co-input/co-partnership (Chan, under review) is crucial.

By adopting this framework, educational institutions can align actions with their policy, ensuring responsible and ethical AI usage while maximizing potential benefits. However, more research is necessary to fully comprehend the potential advantages and risks associated with AI in academic settings. The untapped potential of AI to revolutionize learning still requires further exploration, and conversations surrounding personalization and ethical dilemmas in education are ambiguous, a comprehensive policy requires stakeholders for example, to understand the definition of cheating in the AI era, and how holistic competencies can be developed (if possible) within the AI settings, there are still so much to be unearthed. Merely advocating for AI implementation in education is insufficient; stakeholders need to carefully evaluate which AI technologies to employ, determine the best methods for their use, and understand their true capabilities.

**Declarations:**

The datasets used and/or analysed during the current study are available from the corresponding author on reasonable request

The author declares that one has no competing interests